\documentstyle{mn}

\newif\ifAMStwofonts

\ifoldfss
  \ifCUPmtlplainloaded \else
    \NewTextAlphabet{textbfit} {cmbxti10} {}
    \NewTextAlphabet{textbfss} {cmssbx10} {}
    \NewMathAlphabet{mathbfit} {cmbxti10} {} 
    \NewMathAlphabet{mathbfss} {cmssbx10} {} 
  \fi
  \ifAMStwofonts
    \ifCUPmtlplainloaded \else
      \NewSymbolFont{upmath} {eurm10}
      \NewSymbolFont{AMSa} {msam10}
      \NewMathSymbol{\upi}     {0}{upmath}{19}
      \NewMathSymbol{\umu}     {0}{upmath}{16}
      \NewMathSymbol{\upartial}{0}{upmath}{40}
      \NewMathSymbol{\leqslant}{3}{AMSa}{36}
      \NewMathSymbol{\geqslant}{3}{AMSa}{3E}

    \fi
  \fi
\fi 

\ifnfssone
  \newmathalphabet{\mathit}
  \addtoversion{normal}{\mathit}{cmr}{m}{it}
  \addtoversion{bold}{\mathit}{cmr}{bx}{it}
  \newmathalphabet{\mathbfit} 
  \addtoversion{normal}{\mathbfit}{cmr}{bx}{it}
  \addtoversion{bold}{\mathbfit}{cmr}{bx}{it}
  \newmathalphabet{\mathbfss} 
  \addtoversion{normal}{\mathbfss}{cmss}{bx}{n}
  \addtoversion{bold}{\mathbfss}{cmss}{bx}{n}
  \ifAMStwofonts
    \ifCUPmtlplainloaded \else
      \UseAMStwoboldmath
      \makeatletter
      \new@mathgroup\upmath@group
      \define@mathgroup\mv@normal\upmath@group{eur}{m}{n}
      \define@mathgroup\mv@bold\upmath@group{eur}{b}{n}
      \edef\UPM{\hexnumber\upmath@group}
      \new@mathgroup\amsa@group
      \define@mathgroup\mv@normal\amsa@group{msa}{m}{n}
      \define@mathgroup\mv@bold\amsa@group{msa}{m}{n}
      \edef\AMSa{\hexnumber\amsa@group}
      \makeatother
      \mathchardef\upi="0\UPM19
      \mathchardef\umu="0\UPM16
      \mathchardef\upartial="0\UPM40
      \mathchardef\leqslant="3\AMSa36
      \mathchardef\geqslant="3\AMSa3E
    \fi
  \fi
\fi 

\ifnfsstwo
  \DeclareMathAlphabet{\mathbfit}{OT1}{cmr}{bx}{it}
  \SetMathAlphabet\mathbfit{bold}{OT1}{cmr}{bx}{it}
  \DeclareMathAlphabet{\mathbfss}{OT1}{cmss}{bx}{n}
  \SetMathAlphabet\mathbfss{bold}{OT1}{cmss}{bx}{n}
  \ifAMStwofonts
    \ifCUPmtlplainloaded \else
      \DeclareSymbolFont{UPM}{U}{eur}{m}{n}
      \SetSymbolFont{UPM}{bold}{U}{eur}{b}{n}
      \DeclareSymbolFont{AMSa}{U}{msa}{m}{n}
      \DeclareMathSymbol{\upi}{0}{UPM}{"19}
      \DeclareMathSymbol{\umu}{0}{UPM}{"16}
      \DeclareMathSymbol{\upartial}{0}{UPM}{"40}
      \DeclareMathSymbol{\leqslant}{3}{AMSa}{"36}
      \DeclareMathSymbol{\geqslant}{3}{AMSa}{"3E}
    \fi
  \fi
\fi 

\ifCUPmtlplainloaded \else
  \ifAMStwofonts \else 
    \def\upi{\pi}
    \def\umu{\mu}
    \def\upartial{\partial}
  \fi
\fi

\title{A TiO study of the black-hole binary GRO~J0422+32 in a very low state}
\author[N. A. Webb et al]
       {N.A. Webb,  $^{1}$ \thanks{email:naw@astro.keele.ac.uk}  T. Naylor, $^{1}$  Z. Ioannou, $^{1}$  P.A. Charles, $^{2}$  T. Shahbaz. $^{2}$  \\
 $^{1}$ Department of Physics, Keele University, Keele, Staffordshire, ST5 5BG \\
 $^{2}$ Department of Astrophysics, Nuclear Physics Laboratory, Keble Road, Oxford, OX1  3RH}

\pubyear{1999}

\begin{document}

\maketitle

\begin{abstract}
We present 53 simultaneous photometric (I band) and spectroscopic
(6900-9500\AA) observations of J0422+32, taken during December 1997.
From these we determine that J0422+32 was in its lowest state yet
observed, at I=20.44$\pm$0.08.  Using relative spectrophotometry, we
show that it is possible to correct very accurately for telluric
absorption.  Following this, we use the TiO bands at 7055\AA\ and
7589\AA\ for a radial velocity study and thereby obtain a
semi-amplitude of 378$\pm$16~km~s$^{-1}$, which yields {\it
f(M)}=1.191$\pm$0.021M$_\odot$ and {\it q}=9.0$^{\scriptscriptstyle
+2.2}_{\scriptscriptstyle -2.7}$, consistent with previous
observations.  We further demonstrate that this little explored method
is very powerful for such systems.  We also determine a new orbital
ephemeris of HJD=2450274.4156$\pm$0.0009 + 0.2121600$\pm$0.0000002
$\times$ E.

We see some evidence for an ellipsoidal modulation, from which we
determine the orbital inclination of J0422+32 to be less than
45${^\circ}$.  We therefore calculate a minimum mass for the primary
of 2.22M$_\odot$, consistent with a black hole, but not necessarily
the super-massive one proposed by Beekman et al (1997).  We obtain an
M4-5 spectral type for the secondary star and determine that the
secondary contributes 38$\pm$2\% of the flux that we observe from
J0422+32 over the range 6950-8400\AA.  From this we calculate the
distance to the system to be 1.39$\pm$0.15kpc.

\end{abstract}


\section{Introduction}

From the study of X-ray transient systems, it is possible to determine
the masses of stellar-mass black-holes, by observing the cool
secondary star during quiescence (see e.g. Charles, 1999). In this
paper we present new observations of the late-type secondary star in
the soft X-ray transient GRO J0422+32 (Nova~Per~1992/ V518~Per).
Since the discovery of J0422+32 on August 5th 1992, whilst in
outburst, by the `Compton Gamma Ray Observatory' (Paciesas et al,
1992), there have been three subsequent `mini' outbursts: December
1992 (Harmon, Fishman \& Paciesas, 1992), August 1993 (Filipenko \&
Matheson, 1993) and December-January 1993/4 (Zhao et al, 1993).  The
system has been observed at I=20.03 \cite{Oros95} and I=20.22
\cite{Casa95}, but never, as we show in this paper, in absolute
quiescence.  Therefore, observations in the optical have been dominated
by the flux emitted from the accretion disc around the compact object,
making observations of the M-star difficult.  Here we present results
showing that J0422+32 was fainter still in December 1997, thus
observations of the secondary were more accessible, due to less
contamination from the disc.

Previously, Beekman et al (1997) have determined a minimum mass of the
compact object (black hole) of 15M$_\odot$, from an I band
light-curve.  From calculations by King, Kolb \& Burderi (1996) based
on the `disc instability model', where the mass transfer rate must be
below a critical level  in order for low mass X-ray binary systems to
become transient, a minimum mass of the compact object can be
calculated.  If their assumptions are correct, Beekman et al (1997)
calculated a minimum mass of the compact object of 28M$_\odot$.  This
mass is impossibly large for stellar evolution models. From studying
the ellipsoidal modulation alone, we constrain the black-hole to have
a much lower minimum mass, which is consistent with the evolution
of massive stars that form black holes.

We present a new method for conducting a radial velocity study in such
systems with a late-type secondary star.  Previously, the strong TiO
features seen in M-stars have not been used, as it is very difficult
to correct accurately for the telluric absorption that lies on top of
them.  Instead, the Na~I doublet at 8183, 8195\AA\ is commonly
used.  However, by placing two stars along the slit, J0422+32 and a
local early-type star, we can correct for the telluric absorption very
accurately.  It was also thought that the TiO bands were far too broad
to allow accurate cross correlation.  The bands themselves are broad,
but have within them, very sharp features.  In this work we show, using
the two TiO bands at 7055\AA\ and 7589\AA, how effective
cross-correlating these regions can be.  We also exploit the good
telluric absorption correction to determine the spectral type of the
M-star accurately, through measuring the depths of these two TiO
bands.

A 5.1 hour period, assumed to be the orbital period, has been observed
in most earlier data sets (e.g. Chevalier \& Ilovaisky, 1994; Casares
et al, 1995; Garcia et al, 1996).  Earlier still, values close to 5.1
hours were observed as the system declined from the initial outburst
(e.g. Chevalier \& Ilovaisky, 1992; Kato, Mineshige \& Hirata,
1992). This period has not been observed in other datasets,
e.g. Shrader et al (1994) and Martin et al (1995). We have used our
data to refine the orbital period and to determine a new ephemeris.

\section{Observations}

Three nights of simultaneous spectroscopy and photometry were taken on
1997 December 1-3.  However, most of the third night was lost to bad
weather and time was also lost in the preceding two nights due to
technical problems.  The photometry was taken with the 2.5m Isaac
Newton Telescope (INT), using a Johnson I band filter.  Exposures of 600
seconds were recorded on the `Wide Field Camera'.  The observations
were taken with J0422+32 at 6 different positions on the chip, so that
fringing effects could be corrected later.

The spectroscopy was taken using the 4.2m William Herschel Telescope
(WHT) and the ISIS double beam spectrograph at Cassegrain focus. We
used a slit width of 1 arc-second. Again integration times of 600
seconds were used, so that the orbital smearing was less than the
resolution of the data. On the red arm, the 1124x1124 pixel TEK2 CCD
camera gave a dispersion of 2.7\AA/pixel and a resolution of about
5.4\AA. Copper-neon arc spectra were taken for wavelength calibration.
On the blue arm the EEV10 2148$\times$4200 CCD camera was employed,
giving a dispersion of 0.96\AA/pixel and a resolution of about 3.4\AA,
with a useful wavelength range of $\sim$3000-6000\AA.  Flat-fields
were taken after every image to overcome fringing problems.  A second
star was also placed accurately on the slit to correct for telluric
absorption lines appearing in the spectra of J0422+32.

To ensure that the spectroscopy and photometry were simultaneous,
observations were made every fifteen minutes, on the quarter
hour.  The extra five minutes gave enough time at both telescopes for
the chips to be read out and at the WHT, for arc spectra and tungsten
flats to be taken and for the pointing of the INT to be changed slightly.

\section{Data reduction}

\subsection{Spectroscopy}
\label{sec:spect}

A bias frame was subtracted from each flat-field and target frame.
Each flat-field frame taken closest in time and at roughly the same
hour angle and declination was used to correct the corresponding
spectrum of J0422+32.  To create the flat-field, a polynomial was
fitted to each flat-field, by which it was then divided.  To extract
the spectra, an optimal extraction algorithm \cite{Horn86} was used.

The spectra were wavelength calibrated using the arc spectra taken at
the same position as the object frame.  The J0422+32 spectrum was then
divided by the other star on the slit to remove atmospheric bands.
The second star on the slit was determined to be a late F-star from
the blue spectrum, primarily from the 4000\AA\ break and the calcium H
and K lines.  No further analysis was made of the blue spectra, as the
data were far too noisy.  The red arm spectrum in the region of
interest was intrinsically featureless.  In total there were 53 good
red arm spectra, each with a signal-to-noise of about 3 per pixel.
The second star on the slit was approximately 6 times brighter than
J0422+32, hence there was no degradation in signal-to-noise due to
dividing by the F-star.

\subsection{Photometry}

A median stacked bias frame was subtracted from each frame. However,
the frames had to be flat-fielded in a similar manner to infrared
data, due to a severe fringing pattern on the CCD. This was done by
using sets of six image frames, taken offset from each other, removing
one (hereafter the `single' frame) and leaving five from which a
median stack was made.  The modal value of the median stack was scaled
to be the same as that of the single frame and then the scaled median
stack was subtracted from the single frame of the set.  In this way,
the fringing pattern on the single frame was removed (e.g. Beekman et
al, 1997).  The frames were not flat-fielded, as the flat-fielded
frames were $\sim$30\% noisier, pixel-to-pixel, than the
unflat-fielded frames.

The raw, sky-subtracted counts were obtained for each star using an
optimal extraction technique \cite{Nayl98}.  A nearby star was used to
define a point spread function and this PSF was then used to calculate
a weight map for each pixel of J0422+32 and other stars in the frame.
A weight map clipping radius of almost double the FWHM seeing was used
to recover between 95 and 99 percent of the available signal to noise.
The other star that was taken on the slit during the spectroscopy was
identified from the position angle of the slit on the sky and then
J0422+32 was divided by this to remove sky transparency variations.

The standard stars PG1047 and its associated companions \cite{Land92}
were also observed during the run.  These were reduced in the same manner
as the target and the zero-point from each star was calculated.  A
mean was made of the four zero-points, to find the approximate
zero-point for the CCD.  Using this value, the approximate I-band
magnitude for J0422+32 of I$\simeq$20.4 was calculated, using a frame
with a similar airmass to that of the standard star frame.  This was
further refined, as described in Section~\ref{sec:photan1}.

\section{Data Analysis}
\label{sec:datan}

\subsection{Spectroscopy}
\label{sec:spectan}

To derive a radial velocity curve, various sections of the spectra
were cross correlated against each other.  To do this we used the
cross-correlation technique outlined in Tonry and Davis (1979), to
produce a cross-correlation function (CCF).  A quadratic was fitted
through the CCF peak so as to determine the maximum point,
corresponding to the radial velocity shift for that spectrum.
Initially we cross-correlated individual spectra against each other to
determine the radial velocity variation and to check the period.  We
cross-correlated the region across the TiO band at 7055\AA\ and the
Ca~II region between 8450 and 8750\AA, the latter of which has been
used previously e.g. Casares et al (1995).  A period search was
performed using a Fourier transform routine.  From this and the
`cleaned spectrum', we found no significant periods, including the
orbital period.

The 53 spectra were then binned into 8 phase bins, using the same
ephemeris as Beekman et al (1997), derived from the extensive
photometry of Chevalier \& Ilovaisky (1996) and spectroscopy of
Filipenko, Matheson \& Ho (1995).

\subsubsection{Radial velocities from TiO}

\begin{figure}
\vspace*{12.5cm}         

\includegraphics{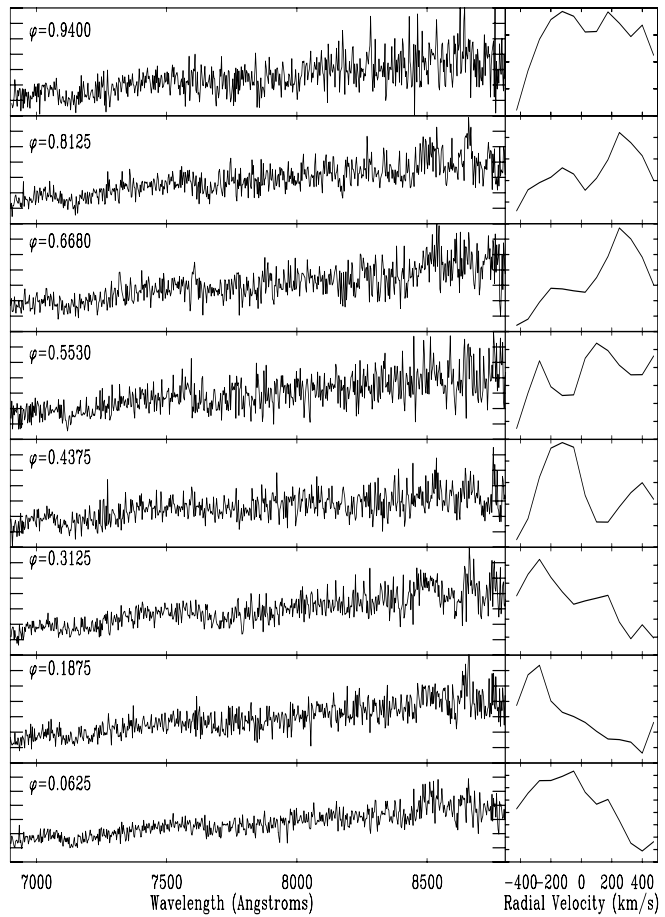}

\caption{In the left hand column, the eight phase-binned spectra and
the average phase at which they were taken. In the right hand column
are the CCFs resulting from cross-correlating the region 7000-7500\AA\
with the weighted mean spectrum of the remaining spectra.}

\label{fig:binsandxcc}
\end{figure}

\begin{figure}
\vspace*{6.5cm}

\includegraphics{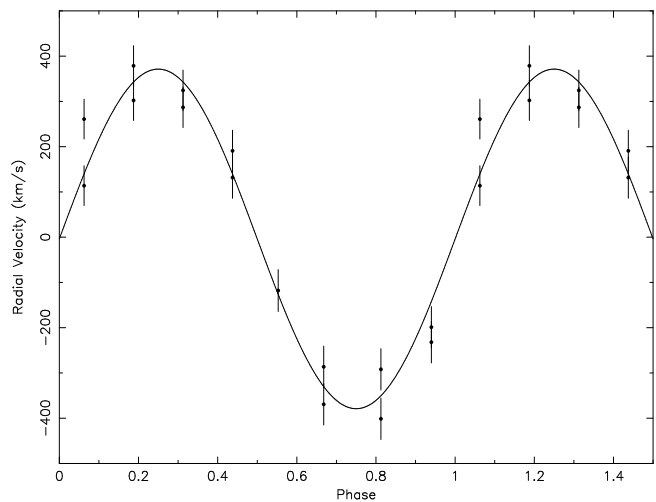}

\caption{Radial velocity curve of J0422+32.  The data have been binned
into 8 bins of size 0.125 in phase.  These results come from cross
correlating the two TiO bands at 7055 and 7589\AA\ against the
weighted mean total spectrum.  One point has not been included in the
fit (from phase 0.55), as there was no clear peak in the CCF.}

\label{fig:rv}
\end{figure}

Each spectrum then had a signal-to-noise of about 7/pixel and was
cross-correlated with the weighted mean of the other spectra (that had
been shifted to rest using a semi-amplitude of 375km s$^{-1}$
(mid-point of generally accepted semi-amplitudes) and the T$_\circ$ of
Beekman et al, 1997, where T$_\circ$ is the first inferior conjunction
of the late-type star for the dataset).  The regions 7000-7500\AA\ and
7500-8000\AA\ were used to cross-correlate the two strongest TiO
bands.  Using the new T$_\circ$ and semi-amplitude obtained from the
radial velocity curve (e.g. Fig.~\ref{fig:rv}), we repeated the above
method until the change in T$_\circ$ between successive calculations
was negligible.  This gave us a T$_\circ$=2450783.6463$\pm$0.0004
(from the beginning of our dataset), a change of 0.013 in orbital
phase (0.003 days) and a new radial velocity of 378$\pm$16
km~s$^{-1}$.  From shifting on this new T$_\circ$, the steps in the
TiO bands and the potassium lines are sharper, indicating that this
T$_\circ$ is more accurate.  The 1$\sigma$ error in the semi-amplitude
was calculated by creating radial velocity errors (weighted according
to the number of spectra in each phase bin) to produce a
${\chi^{\scriptscriptstyle 2}_{\scriptscriptstyle \nu}}$ of
1\footnote{We also tried cross-correlating each phase bin against
other M star spectra (see Sec.~\ref{sec:spectype}), but the results
were not as accurate.}.  Fig.~\ref{fig:binsandxcc} shows the 8
phase-binned spectra and the CCF resulting from cross-correlating the
region 7000-7500\AA\ with the weighted mean spectrum of the remaining
spectra.  The peak of the CCF can clearly be picked out from all but
one phase and the orbital motion of the system can be seen.

To compare our accuracy with previous results, we rescaled the error
bars of Harlaftis et al (1999), so that a fit to their radial velocity
curve also gave a ${\chi^{\scriptscriptstyle 2}_{\scriptscriptstyle
\nu}}$=1.  We obtained a radial velocity of 378$\pm$15 km~s$^{-1}$,
from their data.  The 4\% error on our radial velocity is comparable
to that obtained with 20\% more observing time and using a 10m
telescope (Harlaftis et al, 1999). Thus we have shown that using TiO
bands for radial velocity studies can be a very powerful tool.

\subsubsection{The new ephemeris}
\label{sec:newephem}

From the new T$_\circ$, we re-determined the orbital period, using our
data and that of Chevalier \& Ilovaisky (1996), Filipenko, Matheson \&
Ho (1995), Harlaftis et al (1999) and Casares et al (1995), so as to
extend the base line.  Each T$_\circ$ was adjusted to the phase
convention used here in equation \ref{eq:newephem}.  A period of
0.2121600$\pm$0.0000002d was found to fit all the data, with a
T$_\circ$ of 2450274.4156$\pm$0.0009.  This gives a new ephemeris of,

\begin{equation}
HJD =  2450274.4156(9) + 0.2121600(2) \times E
\label{eq:newephem}
\end{equation}

where HJD is the Heliocentric Julian Date and E the cycle number.
Phase zero for this ephemeris is taken to be the inferior conjunction
of the late-type star.  

\subsection{Photometry}
\label{sec:photan}

\subsubsection{How bright is J0422+32?}
\label{sec:photan1}

From our initial estimate of the magnitude of J0422+32, it appeared
that J0422+32 had become considerably fainter over the 3 years since
it was last observed.  Previously, J0422+32 was found to be $\sim$0.14
magnitudes brighter when observed by Casares et al \shortcite{Casa95}
in December 1994 and I$\sim$0.33 magnitudes brighter when observed by
Orosz \& Bailyn \shortcite{Oros95} in October 1994.  To verify this,
we used the following method. 

The Casares et al \shortcite{Casa95} observation is the lowest
published value to date.  We reanalysed the Casares et al
\shortcite{Casa95} data by obtaining the frames taken at the JKT in
December 1994 from the RGO Astronomy Data Centre.  These we reduced in
the same manner as the J0422+32 photometry that we obtained in
December 1997 from the INT, although a simple flat-fielding technique
was used, as there was no fringing effect on these CCD frames.
Photometry was carried out on the 1994 data of J0422+32 and 7 stars in
the neighbourhood and on the same stars in the 1997 data.  Both the
frames used were from a similar phase (${\phi\simeq}$0.67).  In each
case, J0422+32 was divided by each of the other 7 stars in the frame
and the relative counts compared from the two years.  The 1997 data
was found to be 22$\pm$3\% fainter than the 1994 data of
I=20.22$\pm$0.07 (where the 3\% error is the standard deviation
calculated from the ratio between the 1997 and 1994 data for each of
the 7 stars). This indicates that we observed J0422+32 at an I
magnitude of 20.44$\pm$0.08.  This is the lowest state that J0422+32
has been observed to date, which indicates that in 1994, it had not
yet reached its quiescent value, following the outburst and subsequent
mini-outbursts.

\subsubsection{Orbital modulation}
\label{sec:photan2}

To search for periodicities in the photometry, a Fourier transform
routine was used.  From the power spectrum, we found no significant
evidence for the orbital period reported by others (e.g. Filippenko,
Matheson \& Ho, 1995), nor any other period.

However, as binning the spectroscopic data on the 5.1 hour period had
been so successful, the data were then binned into the same 8 phase
bins, of 0.125 of the orbital period.  Each bin had a minimum of 5
points in it.  The weighted mean relative flux was calculated and an
average phase, for each bin.  Plotting relative flux against phase, an
ellipsoidal modulation, as seen by Beekman et al (1997), Chevalier \&
Ilovaisky (1996), Martin et al (1995) etc is therefore evident.
However, we observe excess flux between phases $\sim$0.4-0.6, which
may be due to the irradiation of the M star by the inner edge of the
accretion disc (e.g. Shahbaz, 1994), although why the X-ray heating
should increase during this low state, is unclear.  From examining the
database from the All-Sky Monitor on the X-ray Timing Explorer (XTE)
for J0422+32 around this observation period, the X-ray activity was
minimal.  It is however possible, that since J0422+32 had declined
from outburst in December 1997, the disc opening angle should also
have decreased (see e.g. Webb et al, 1999).  Therefore more X-rays
from the inner parts of the disc would be able to impinge on the
secondary star, and thus heat the surface between phases 0.4-0.6.

Our interest is to measure the maximum allowed amplitude in this
light-curve, to constrain the inclination of this system (see
Sec.~\ref{sec:orbpar}).  Clearly, a larger amplitude is possible if we
ignore the `X-ray heated' points.  We therefore fitted the flux
variation between phases $\sim$0.6-1.4.  The maximum flux
variation is 0.014.  Taking into account that only 38\% of the flux
observed emanates from the M star (see Section~\ref{sec:distance}),
the maximum percentage flux variation is 16\% from the secondary star
alone.

\begin{figure}
\vspace*{6cm}         
\includegraphics{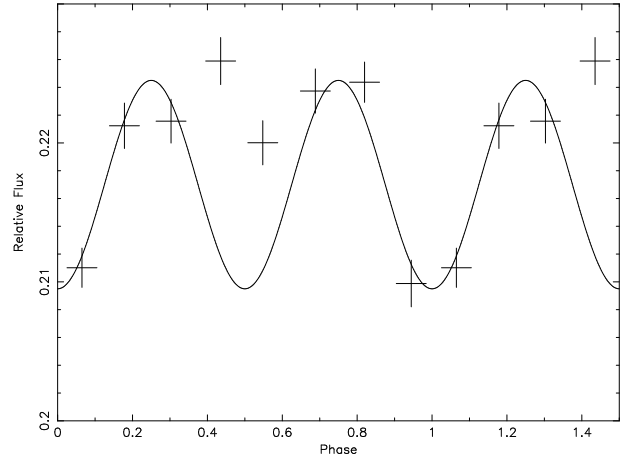}

\caption{Photometric data divided by standard star in the frame and
binned into 8 x 0.125 of a phase bin.  Errors are the standard
deviation of the average time for each bin and the standard error of
the flux for each bin.  Solid line shows the maximum double sine wave
fit, excluding the points at phases 0.44 and 0.55, as these phases may be 
affected by irradiation from the accretion disc.}

\label{fig:doublehump}
\end{figure}

\section{Results and discussion}

\subsection{Flux Calibration}
\label{sec:slitcor}

\begin{figure}
\vspace*{6cm}         

\includegraphics{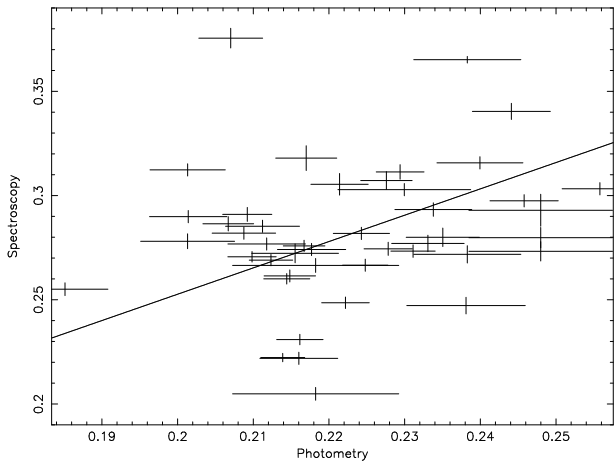}

\caption{Individual relative spectroscopic fluxes plotted against
relative simultaneous photometry and the best fit straight line
through the origin to these.  The need for spectroscopic slit
correction is clearly evident.}

\label{fig:slitcor}
\end{figure}

We plotted the relative flux from each spectroscopic observation
against each simultaneous photometric observation, to compare the
fluxes. This was done by summing the J0422+32 spectra over the range
of the photometric pass-band (7250-9500\AA) and dividing this by the
sum of the comparison star over this same range and then plotting this
against the photometric observations (Fig.~\ref{fig:slitcor}).  There
is a large scatter about the best fit line to the spectroscopy verses
photometry, showing that there are considerable differences in the
light losses at the slit between the two stars.  We therefore decided
to correct for these slit losses, by multiplying each J0422+32
spectrum by the ratio between the photometry and the spectroscopy.

\begin{figure*}
\begin{minipage}{170mm}
\vspace*{10cm}         

\includegraphics{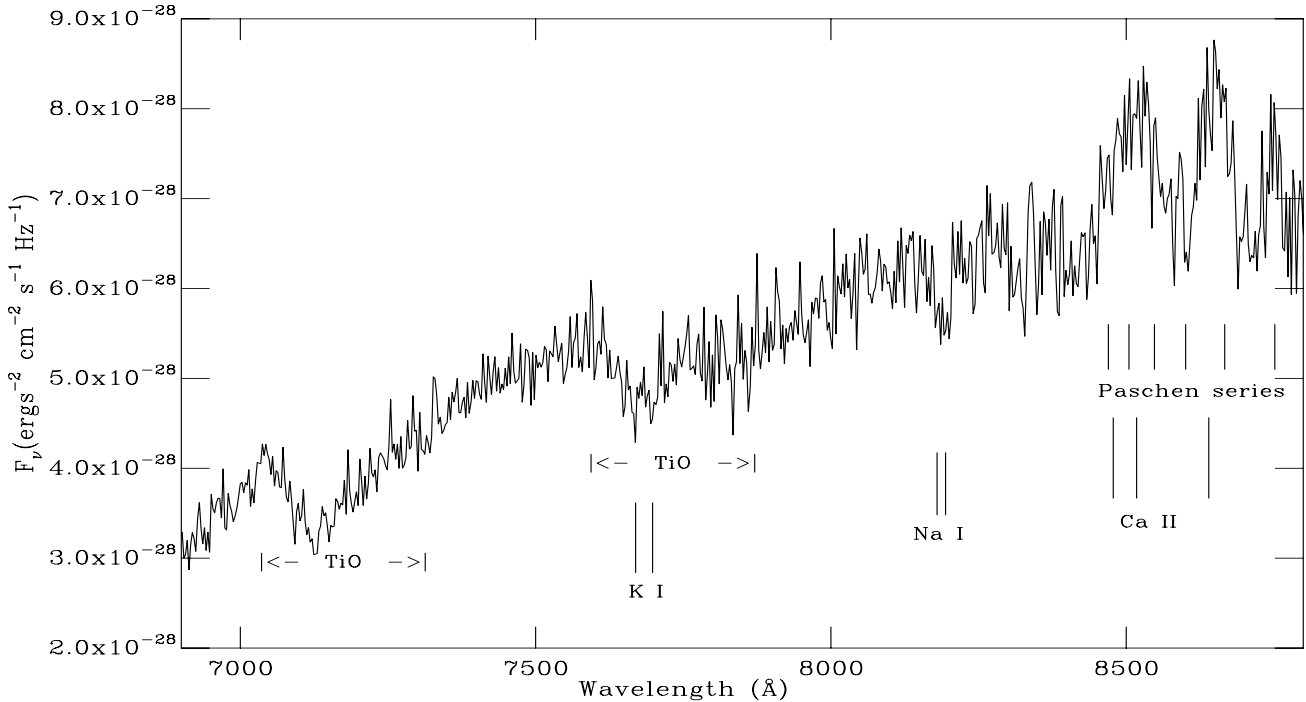}

\caption{The fluxed, dereddened, weighted mean of 53 spectra.  The TiO
bands at 7055 and 7589\AA\ are clearly visible, as are the potassium
lines within this second band.  The sodium doublet at 8183,95\AA\ is
also visible as are Paschen lines at 8470, 8505, 8548, 8601, 8667,
8752\AA, superposed over Ca~II lines at 8498.0, 8542.1 and 8662.1\AA.}

\label{fig:simpleadd}
\end{minipage}
\end{figure*}

The second (local) star observed on the slit to be used as a
comparison star, was flux calibrated using the flux standard
SP~2209+178 (BD~+17~4708), in the following way.  The instrument
response function derived from the flux standard was used to flux the
weighted mean of the 53 spectra of the comparison star.  The flux
level was virtually insensitive to changes in the airmass. A smooth
curve was fitted to the fluxed mean comparison star spectrum.  The
J0422+32 spectrum (divided by the comparison star) could then be
fluxed by multiplying by this curve and remain corrected for
telluric absorption.  The J0422+32 fluxed spectrum was dereddened
using an E(B-V)=0.3 (Harlaftis et al, 1999 and Beekman et al, 1997)
and barycentric radial velocity corrected (Fig.~\ref{fig:simpleadd}).

\subsection{Search for the ellipsoidal modulation}
\label{sec:fluxdef}

We searched for the ellipsoidal modulation of the secondary star in
two ways.  We measured flux deficits for each of the 8 phase bins, as
described in Wade \& Horne (1988), but the errors were large
(25-100\%).  To get a higher signal-to-noise, varying proportions of
the summed spectrum (Fig.~\ref{fig:simpleadd}) were subtracted from
each of the 8 phase binned spectra.  The region 7000-8100\AA\ was
used, as it contains two strong TiO bands, but no Ca lines.  A line
was fitted to the residuals and we used ${\chi^2}$ to determine the
best fitting proportion of the summed spectrum at each phase.  This
measure could also reveal ellipsoidal variation.  Although the errors
were smaller (10-20 percent), only a singly-humped lightcurve was seen
rather than a double-humped curve.  This is probably due to
irradiation of the front face of the M star, where the hotter
temperatures diminish the quantity of TiO.

\subsection{Orbital parameters of J0422+32}
\label{sec:orbpar}

Using the relationship between the amplitude of the ellipsoidal
variation and the inclination of the system (Shahbaz 1994), we
determined the maximum orbital inclination of J0422+32 to be
45$^{\circ}$, for the maximum amplitude obtained with the photometry
(in which we can see evidence of ellipsoidal variation, see
Sec.~\ref{sec:photan2}).  Using the semi-amplitude obtained from our
radial velocity study of 378$\pm$16 km~s$^{-1}$, we obtained a mass
function of 1.191$\pm$0.021M$_\odot$.  We used this new mass function,
the $v$sin$i$ of 90$^{\scriptscriptstyle +22}_{\scriptscriptstyle
-27}$ km s$^{-1}$ \cite{Harl99} and the inclination, to obtain a new
{\it q} value of 9.0$^{\scriptscriptstyle +2.2}_{\scriptscriptstyle
-2.7}$, where {\it q} is the mass of the primary (compact object)
divided by the secondary.  From this we can constrain the mass of the
primary to be greater than 2.2M${_\odot}$.  Although this is below
2.9M$_\odot$ (the upper bound on the mass of a neutron star, Kalogera
\& Baym (1996)), it is well above the canonical value for a neutron
star (1.44M$_\odot$) and would imply a super-massive neutron star if
the compact object were as little as 2.2M${_\odot}$.  Further to that,
from evidence in previous work, e.g. Sunyaev et al (1993), who show
that J0422+32 has many of the same characteristics as seen in the
other black hole candidate X-ray transients, J0422+32 must indeed
contain a black hole.

\subsection{The spectral type of the secondary star in J0422+32}
\label{sec:spectype}

The dereddened summed spectrum was normalised to unity at 7500\AA. We
measured the flux deficits across the two strongest TiO bands, namely
those at 7055\AA\ and 7589\AA, as described in Wade \& Horne (1988)
and Section~\ref{sec:fluxdef}.  Having continuum prior to the band at
7055\AA, unlike Wade \& Horne, we could fit the continuum across this
first band between 6900-7000\AA\ and 7450-7550\AA.  We also measured
the drop between 7140 and 7190\AA.  Over the second band, we again
used the same regions as Wade \& Horne (1988); the continuum was
fitted between 7450-7550\AA\ and 8130-8170\AA\ and the drop measured
between 7640-7690\AA.  From this we obtain the ratio of
d${_\nu}$(${\lambda}$7665)/d${_\nu}$(${\lambda}$7165) to be
0.91$^{\scriptscriptstyle +0.25}_{\scriptscriptstyle -0.17}$. This
implies that the secondary star in J0422+32 is an M4-5 star, using
Wade \& Horne's table 3.

As this result gives a slightly later spectral type for the secondary
star than Harlaftis et al (1999), M2$^{\scriptscriptstyle
+2}_{\scriptscriptstyle -1}$ or Casares et al (1995),
M2$\pm$2~\footnote{NB: Martin et al (1995) propose that the secondary
star is anything between an M0-5}, we employed a second method in
order to confirm the spectral type.  Spectra of Gliese stars Gl 338A
(M0 star), Gl 361 (M2), Gl 436 (M3) and Gl 402 (M5), were observed and
reduced in the same manner as J0422+32.  Although simultaneous
spectrophotometry was not possible, a bright, rapidly rotating B star
was observed at a similar airmass.  By dividing the Gliese star
spectrum by the B star, we corrected for telluric absorption.  The
Gliese stars were fluxed in the same manner as the J0422+32 spectrum
(Section~\ref{sec:slitcor}).  These too were normalised to unity at
7500\AA.

Varying proportions of the normalised Gliese stars were then
subtracted from the normalised, dereddened spectrum of J0422+32.  A
quadratic was fitted to the residuals and the
$\chi^{\scriptscriptstyle 2}_{\scriptscriptstyle \nu}$ of the fit to
this quadratic was calculated as a measure of the goodness of
fit. This also gives an indication of the proportion of the flux
observed that emanates from the secondary star.  From this, the best
fitting spectrum to J0422+32 was spectral type M5.  However, as we had
failed to observe an M4 star, due to the onset of bad weather, we
constructed a mean of the M3 and the M5 star to simulate a star that
was approximately an M4.  This gave as good a fit as the M5
star. Also, as the error on the reddening is possibly large, we
carried out both experiments on the non-reddened spectrum to check
that the uncertainty in the reddening correction had not affected our
spectral typing.  However, both the reddened and non-reddened spectra
gave the same results.  We therefore conclude that the secondary star
in J0422+32 is M4-5.  The reason for the discrepancy between our
measurements and those of previous workers is probably their shortage
of later type spectral standards. Both Harlaftis et al (1999) and
Casares et al (1995) observed no stars later than M2, with the same
apparatus used to observe J0422+32.

\subsection{Comparison with previous mass determination}

Previously Beekman et al (1997) used a method based on the disc
instability model, following work done by King, Kolb \& Burderi
(1996), to determine the minimum mass of the compact object (revised
version, Beekman 1999 and King \& Kolb, 1999).  The mass transfer rate
must be below a critical level for a low mass X-ray binary system to
become transient and therefore a minimum mass can be calculated. From
this Beekman et al (1997) determined that the minimum mass of the
compact object in J0422+32 was at least 32M$_\odot$.  This was
calculated using the mass and radius of a main-sequence M2 star, as in
Gray (1992). However, the values for M stars are uncertain (Jones et
al, 1994).  Using values from Leggett et al (1996) and Jones et al
(1994) for the M2 standard star Gl 411 and the revised calculations as
given in Beekman (1999) and King \& Kolb (1999), we obtain a much more
reasonable minimum mass of the black-hole of 11.9M$_\odot$.  However,
we have shown that the spectral type of the secondary star is somewhat
later than M2.  For an M4.5 star (Gl 268), a star that is consistent with
our spectral classification of M4-5, we calculate that the black hole
mass must be greater than 0.75M$_\odot$. Although this seems
considerably smaller than our minimum mass derived from the radial
velocity curve, it should be noted that the revised minimum compact
object masses in Beekman (1999) and King \& Kolb (1999) are all
considerably lower than the minimum masses calculated from the
respective radial velocity curves.  King \& Kolb (1999) also describe a
second method by which they determine the minimum mass of the compact
object, from the spectral type and therefore the mass of the secondary
star, for stars of spectral type K and earlier.  However, as we have
shown that the secondary star in J0422+32 is of spectral type M4-5, this
method is not applicable to this system.  We have therefore solved the
problem of the black hole in J0422+32 having an impossibly large mass
and also demonstrated that these equations support our radial velocity
calculations.

\subsection{Distance}
\label{sec:distance}

We calculated that 38$\pm$2\% (2$\sigma$ error) of
the light emanates from the secondary star in the spectral range
6950-8400\AA, by subtracting varying proportions of the normalised
Gliese stars from the normalised spectrum of J0422+32
(Sec.~\ref{sec:spectype}).  This implies that 62$\pm$2\% of the
observed flux originates from the accretion disc in this region.  This
is considerably more, although still consistent with Casares et al
(1995), who estimated that the secondary is emitting 48$\pm$8\% of the
light in the spectral range 6700-7500\AA.

From an E(B-V)=0.3 and A$_v$/E(B-V) = 3.1 (Seaton, 1979 and Howarth,
1983), we obtain an A$_v$ of 0.93.  The I band absolute magnitude of
an M4 star, Gl 213 \cite{Kirk93} is M$_I$=9.86, and with only 38\% of
the flux coming from the secondary, it would have an I band magnitude
of $\simeq$21.50.  Hence, we calculate that the distance to J0422+32
is 1.39$\pm$0.15 kpc.  This is not only consistent with, but also far
more accurate than other previous measurements of distance,
e.g. Callanan et al (1995), who give a lower limit of 1kpc and a
probable upper limit of 2kpc.

Using this value for the distance and the X-ray flux of
3.5$\times$10$^{-8}$ ergs cm$^{-2}$ s$^{-1}$ (Harmon et al, 1992) in
the 40-230 keV range during the main outburst, we calculate that the
luminosity is somewhat lower than the Eddington luminosity
(2.8$\times$10$^{38}$ ergs s$^{-1}$) at only 8$\times$10$^{36}$ ergs
s$^{-1}$.  There may, however, be considerable flux in other
wavebands, for which there are no data available (Shahbaz, Charles \&
King, 1998).
 
\section{Conclusion}

From our data, we determine that J0422+32 was in its lowest state yet
observed during December 1997, at I=20.44$\pm$0.08.  Following a
comprehensive search we measure a period of 0.2121600$\pm$0.0000002 days,
consistent with previous analysis.  From the same radial velocity
study, we determine a tightly constrained semi-amplitude of
378$\pm$16 km s$^{-1}$.  Thus we have shown that using TiO bands
for radial velocity studies is a very powerful tool.  We also
determine a new T$_\circ$ of 2450274.4156$\pm$0.0009.

Further to this we see some evidence for ellipsoidal modulation,
implying a maximum inclination for J0422+32 of 45$^\circ$.  We obtain
a new mass function of 1.191$\pm$0.021 M$_\odot$ and {\it
q}=9.0$^{\scriptscriptstyle +2.2}_{\scriptscriptstyle -2.7}$, from
which we have calculated a minimum mass for the primary of
2.22M${_\odot}$, consistent with a black-hole, but not necessarily as
a super massive one.  However, this is only a minimum mass and the
maximum is not constrained.  We also revise the calculations of
Beekman et al (1997) to demonstrate that the minimum possible mass of
the black hole is similar to other black hole systems.

We have shown that it is important to perform a slit correction to the
spectra, to calculate the correct flux for such a faint
system.  We obtain a spectral type of the secondary of M4-5 and
determine that the secondary star contributes 38$\pm$2\% of the flux
that we observe from J0422.  From this we calculate the distance to
the system to be 1.39$\pm$0.15 kpc.

\section*{Acknowledgments}

We are grateful to the RGO Astronomy Data Centre for providing us with
JKT data taken in December of J0422+32.  Thanks also go to R. Hynes
for many useful comments on this work.  NAW is supported by a PPARC
studentship.  TN is in receipt of a PPARC Advanced Fellowship.

\end{document}